\documentclass[]{aa} 

\usepackage{txfonts}
\usepackage{longtable}  
\usepackage{rotating}
\usepackage{natbib}
\usepackage{graphics}
\usepackage{graphicx}
\usepackage{bmpsize}
\usepackage{psfrag}
\usepackage{amssymb}
\usepackage{multicol}
\usepackage{pdflscape}
\usepackage{rotating}
\bibpunct{(}{)}{;}{a}{}{,}
\begin{document}

\title{A new catalogue of Str{\"o}mgren-Crawford $uvby\beta$ photometry}

\author{E.~Paunzen}
\institute{Department of Theoretical Physics and Astrophysics, Masaryk University,
Kotl\'a\v{r}sk\'a 2, 611\,37 Brno, Czech Republic \\
\email{epaunzen@physics.muni.cz}}

   \date{} 
  \abstract
{The $uvby\beta$ photometric system is widely used for the study of various Galactic and
extragalactic objects. It measures the colour due to temperature differences,
the Balmer discontinuity, and blanketing absorption due to metals.}  
{A new all-sky catalogue of all available $uvby\beta$ measurements from the
literature was generated.}  
{The data for the individual stars were cross-checked on the 
basis of the Tycho-2 catalogue. This catalogue includes very precise celestial coordinates, but
is magnitude and spatial resolution limited. However, the loss of objects is only 
marginal and is compensated for by the gain of homogeneity.}  
{In total, 298\,639 measurements
of 60\,668 stars were used to derive unweighted mean indices and their errors. Photoelectric
and CCD observations were treated in the same way.}  
{The presented data set can be used for various applications such as new calibrations of astrophysical parameters, the
standardization of new observations, and as additional information for ongoing and forthcoming all-sky surveys.}

\titlerunning{}
\authorrunning{}
\maketitle

\section{Introduction}\label{introduction}
One of the most successful and powerful astronomical photometric systems is  $uvby\beta$ 
introduced by Bengt Str{\"o}mgren \citep{Stro56,Stro66} and extended by David Crawford \citep{Craw58}.
It was mainly designed to investigate stars and their basic astrophysical characteristics in 
an acurate way. It measures the effective temperature, the Balmer discontinuity, and blanketing due to metallic lines.
Furthermore, it can be used to estimate the interstellar extinction and reddening.
Several reddening-free indices for many different purposes and spectral type regions have been developed so far. In addition, some
very successful extensions have been developed such as the $\Delta a$ \citep{Paun05} and Stromvil
\citep{Stra96, Stra99} systems. 

More than two dozen papers have been dedicated to the calibration of stellar
astrophysical parameters of objects across the complete Hertzsprung-Russell diagram 
\citep[HRD; e.g.][]{Olse88,Gray92}. 

It was also used, for example,  to investigate both open and globular clusters \citep{Beav13,Cala14},
the stellar population in the Magellanic Clouds \citep{Liva13}, and the ages of early-type 
galaxies \citep{Rako08}. These are just a few of the applications of this photometric system.
Recently, \citet{Wang14} announced a new all-sky $uvby\beta$ survey to a completeness limit of about 19th magnitude
on the basis of one-metre class telescopes.

In order to prepare  the grounds for new astrophysical calibrations, and also to supplement spectroscopic,
photometric, and kinematic surveys such as Gaia \citep{Jofr14}, LAMOST \citep{Li15}, and RAVE \citep{Piff14}, 
it is essential to have a homogeneous set of mean $uvby\beta$ photometry. Starting with \citet{Lind73}, a continuous update
of the available photoelectric measurements were compiled until the last version by \citet{Hauc98} at the  
Institut d'Astronomie de l'Universit{\'e} de Lausanne (Switzerland). Since then, no efforts have been made to also
include CCD data into this compilation.

In this paper, a new catalogue of the published and available $uvby\beta$ measurements, through to the end of 2014, is presented.
It builds upon the data set published by \citet{Hauc98}. It includes photoelectric as well as CCD data.
The cross reference of the objects has been done using the positional data of the Tycho-2 catalogue \citep{Hog00}. 
Such a procedure is especially important for close binary systems, high proper motion stars, and objects in dense star cluster fields. 
In total, 298\,639 data points were used to get the final mean values for 60\,668 stars with $-$1.09\,$<$\,$V_{\mathrm T}$\,$<$\,+14.05\,mag.
However, about 98\% stars are between 4th and 12th magnitude.

\begin{table}
\caption{Number of individual measurements for each index, the number of objects, and the mean of the standard deviation.}
\label{n_measurements}
\begin{center}
\begin{tabular}{lccc}
\hline
Index           & $N_{\mathrm{meas}}$ & $N_{\mathrm{stars}}$ & $\overline{\sigma}$ \\
                &                    &                     & [mag]               \\
\hline
$b-y$           & 81252 & 55076 & 0.008 \\      
$m_{\mathrm 1}$ & 80164 & 54070 & 0.010 \\
$c_{\mathrm 1}$ & 80087 & 53981 & 0.012 \\
$\beta$         & 57136 & 39917 & 0.011 \\
\hline   
\end{tabular}    
\end{center}                                      
\end{table}

\section{Sample selection and analysis}

The Tycho-2 catalogue \citep{Hog00} and its two supplements were used as the basis to identify and cross-reference the stars and their measurements.
The catalogue provides celestial positions, proper motions, and two-colour photometric data for about 2\,550\,000 stars. Components of binaries 
with separations down to 0.8'' are included. The catalogue is about 90\% complete to magnitudes of $V$\,$\sim$\,11.5\,mag with the faintest
objects around the  15th magnitude. Only objects included in the Tycho-2 catalogue were considered in the following. With that constraint very close
binary systems, fainter objects, and stars in very dense fields (star clusters) are not incorporated, but possible false identifications
are minimized.

The starting point for the compilation of the available $uvby\beta$ data was the catalogue by \citet{Hauc98}, which includes 
only photoelectric measurements.
It contains 105\,873 individual measurements of 63\,313 stars without an a priori magnitude limitation. 
Since then, regular updates until 2007 were incorporated in the
General Catalogue of Photometric Data \citep[GCPD, ][]{Merm97}. However, the most recent version is not available in a compiled form
and does not include any CCD observations. For the new catalogue, the literature though the end of 2014 was searched for photoelectric 
and CCD $uvby\beta$ measurements. 

The weighting procedure and calculation of the means by \citet{Hauc98} was not duplicated here.
They introduced a two-step iterative procedure, where the first step consists of a simple weighted mean, 
the weight being the number of measurements to the 2/3 power and a publication weight.
The latter are individual heuristic weights between zero and four for each publication depending on
the quality of the data (for example, known offsets).
The second step uses the differences of the individual values to the mean to compute the final weighted 
mean values. The main reasons why this procedure was not applied is that a) most authors are no longer listing the actual number of individual
measurements/frames/images and b) there is no objective way to derive the publication weight.
The comparison of the internal accuracy of CCDs to photoelectric measurements and the reliability of
standardization procedures has often been  discussed \citep{Bess05}. In particular,  the found differences in
the blue wavelength region ($u$ filter) make a comprehensive weighting complicated.

A simple averaging, without weighting, of all individual measurements for one single star was applied and the standard
deviation of the mean calculated. For only one available data point either its error from the corresponding reference or no
value is listed.

The unique numbering system of \citet{Hauc98} was not continued. Its designation `uvby98' is followed by eight digits,
which is based on the identification according to the Geneva seven-colour photometric system \citep{Rufe88}. It is based
on a hierarchical system of secondary catalogues and numbering systems. There are special codes for members of clusters,
associations, different kind of stars (for example, Faint Blue Stars, White Dwarfs, and Emission-Line Stars), and common
acronyms. Since the new compilation is based on the Tycho-2 catalogue, those numbers together with celestial coordinates, are
used as identifiers. Several cross identification of the Tycho-2 catalogue with other sources are available in the literature
\citep[e.g. ][]{Fabr02}.

Table \ref{n_measurements} lists the number of individual measurements for the four indices.
In total, 298\,639 data points were used to get the final mean values. This is almost a factor of
three more than in the catalogue by \citet{Hauc98}.

The final catalogue includes the four mean observables $(b-y)$, $m_{\mathrm 1}$, $c_{\mathrm 1}$, and $\beta$, their
standard deviations, the number of individual measurements, and the complete information of the Tycho-2 catalogue. 
 A separate file will also be provided in the VizieR database\footnote{http://vizier.u-strasbg.fr/viz-bin/VizieR} and a 
mirror of the GCPD\footnote{http://gcpd.physics.muni.cz}, which includes all individual measurements and the references.

\begin{figure}
\begin{center}
\includegraphics[width=85mm, clip]{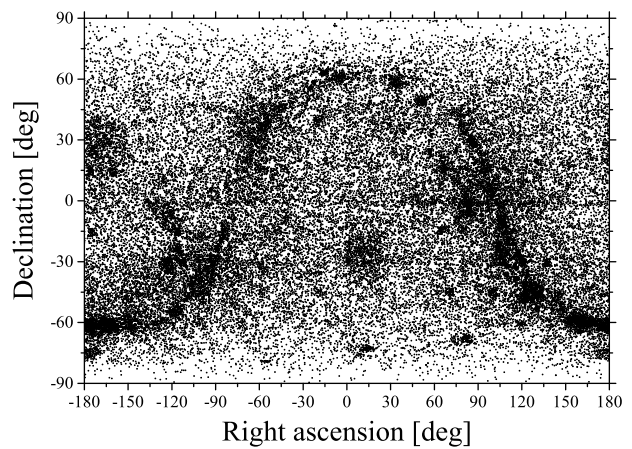}
\caption{The distribution of the 60\,668 catalogue stars on the sky.}
\label{alpha_delta} 
\end{center} 
\end{figure}

\begin{figure}
\begin{center}
\includegraphics[width=85mm,clip]{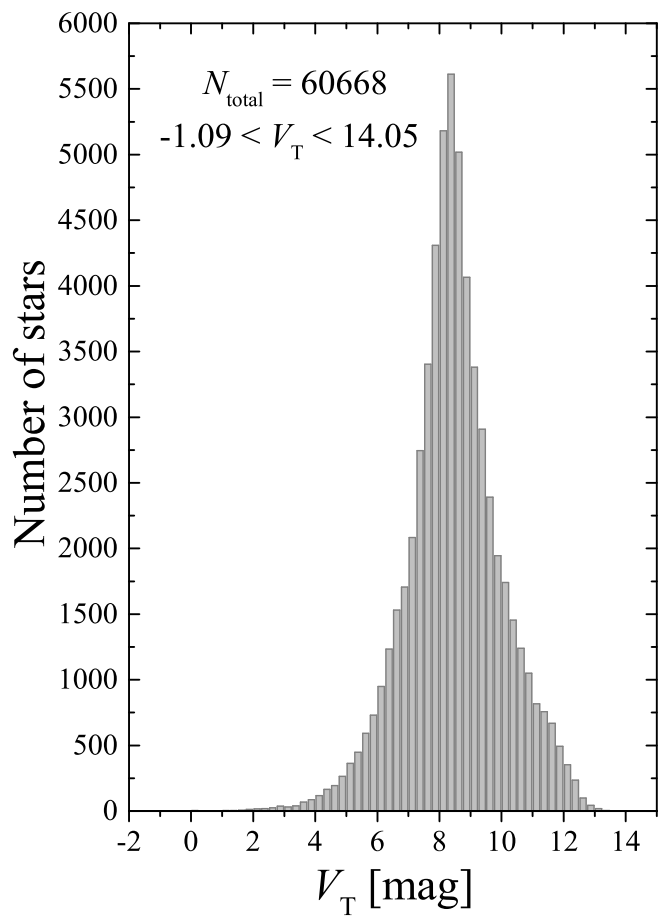}
\caption{The $V_{\mathrm T}$ distribution of the 60\,668 catalogue stars.}
\label{V_stat} 
\end{center} 
\end{figure}

\section{Results and outlook}

The final catalogue represents a sample of stars distributed over the whole sky as shown in Fig. \ref{alpha_delta}.
Several very important Galactic features such as the disk and  the north and south poles are seen. There is a belt
of stars visible  at zero declination,  which is due to many standard stars measured in this region being accessible
from both the northern and southern hemispheres.

The distribution of the $V_{\mathrm T}$ magnitudes of the catalogue stars is shown in Fig. \ref{V_stat}. The
transformation to Johnson $V$ can be done by taking in account a colour term with a factor of 0.09, and an additional 
correction $\delta V$. The latter reaches not more than 0.05\,mag \citep{Khar01}. The catalogue includes stars from which
98\% are in the magnitude range of 4.0\,$<$\,$V_{\mathrm T}$\,$<$\,12.0\,mag, with a peak at about 8.5\,mag.

One of the most severe complications when using photometric indices for calibrating and estimating stellar parameters is
the interstellar extinction. Depending on the Galactic location and the location of dust and gas clouds, it could reach up to 4.0 mag kpc$^{-1}$ \citep{Chen14}.
Therefore, even in the early history of the photometric systems, researchers looked for a feasible work around. While it normally only works
for hot stars in broadband photometric systems, such as Johnson $UBV$ or 2MASS $JHK_{\mathrm{S}}$, \citet{Stro66} already suggested  using the reddening independent indices 
$\left[m_{\mathrm 1}\right]$, $\left[c_{\mathrm 1}\right]$, and $\beta$.

The $\left[m_{\mathrm 1}\right]$ index measures the depression due to metal lines around 4100\AA\, (metallicity),
$\left[c_{\mathrm 1}\right]$ measures the Balmer jump (luminosity), and $\beta$ measures the strength of the H$\beta$ line
(effective temperature). Depending on the spectral region, these indices could also be  slightly sensitive to other parameters than
the listed basic astrophysical ones \citep{Stro66, Gola74}. 

For the calculation of the reddening-free parameters $\left[m_{\mathrm 1}\right]$ and $\left[c_{\mathrm 1}\right]$, the coefficients 
by \citet{Craw76} were used:
\begin{eqnarray}
\left[m_{\mathrm 1}\right] & = & m_{\mathrm 1} + 0.33\left(b-y\right) \\
\left[c_{\mathrm 1}\right] & = & c_{\mathrm 1} - 0.20\left(b-y\right) 
.\end{eqnarray} 

\begin{figure}
\begin{center}
\includegraphics[width=85mm,clip]{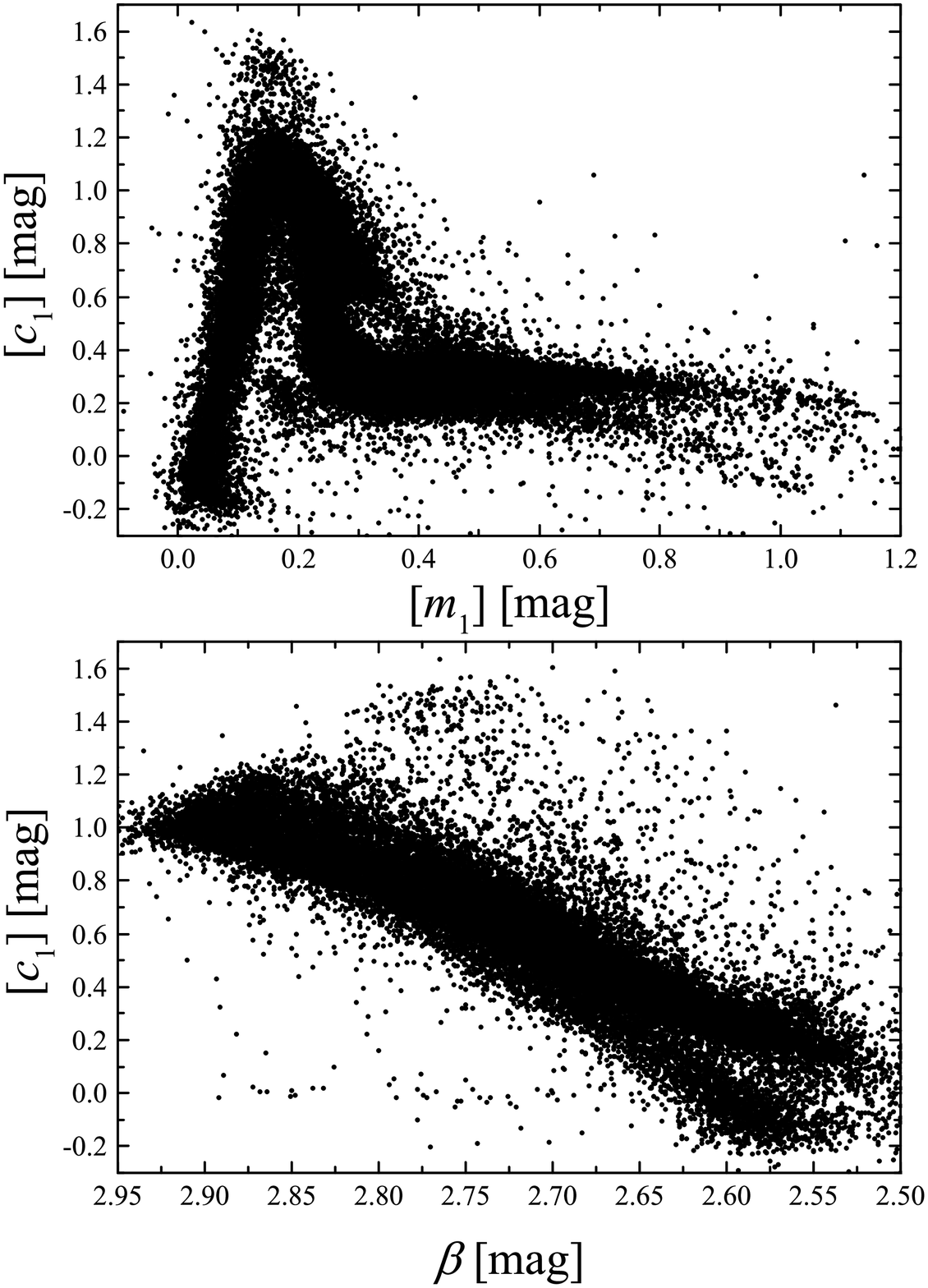}
\caption{The $\left[c_{\mathrm 1}\right]$ versus $\left[m_{\mathrm 1}\right]$ (upper panel) and $\left[c_{\mathrm 1}\right]$ versus $\beta$ (lower panel)
diagrams for the catalogue stars.}
\label{indices} 
\end{center} 
\end{figure}

Figure \ref{indices} shows the $\left[c_{\mathrm 1}\right]$ versus $\left[m_{\mathrm 1}\right]$ and $\left[c_{\mathrm 1}\right]$ versus $\beta$
diagrams for the stars included in the catalogue. In the upper panel the main-sequence band with the bifurcation of the giants
is clearly visible. The hot supergiants are located above the main-sequence at $\left[m_{\mathrm 1}\right]$\,=\,0.2\,mag.
The clump of low-metallicity stars, i.e. intermediate Population I and true Population II objects, is located
around $\left[c_{\mathrm 1}\right]$\,=\,$\left[m_{\mathrm 1}\right]$\,=\,0.2\,mag. The lower panel shows the main-sequence
of stars with the bifurcation of cool giants. As in the former diagram, supergiants are above the main sequence, whereas subdwarfs and binary stars
are below it. 

This new data set can be used  for new calibrations of astrophysical parameters and also for the
standardization of new observations and as additional information for ongoing as well as forthcoming all-sky surveys.

\section*{Acknowledgments}
This project is financed by the SoMoPro II programme (3SGA5916). The research leading
to these results has acquired a financial grant from the People Programme
(Marie Curie action) of the Seventh Framework Programme of the EU according to the REA Grant
Agreement No. 291782. The research is further co-financed by the South-Moravian Region. 
It was also supported by the grant 7AMB14AT015 and
the financial contributions of the Austrian Agency for International 
Cooperation in Education and Research (BG-03/2013 and CZ-09/2014). 
This research has made use of the WEBDA database, operated at the 
Department of Theoretical Physics and Astrophysics of the Masaryk University.
This paper is dedicated to Roswitha M{\"u}ller who died during its preparation.
This work reflects only the authors' views, and the European 
Union is not liable for any use that may be made of the information contained therein.

\label{lastpage}
\end{document}